\documentclass[aps,prd,preprintnumbers,twocolumn]{revtex4}

\usepackage{fancyhdr}
\usepackage{amsmath}
\usepackage{amssymb}
\usepackage{color}
\definecolor{bl}{cmyk}{1,1,0.3,0}
\definecolor{gr}{rgb}{0,0.35,0.1}
\usepackage{hyperref}
\hypersetup{colorlinks=true,bookmarks=true,urlcolor=gr,linkcolor=gr,citecolor=gr}

\newcommand{\dif}{\textrm{d}}

\begin{document}

\pagestyle{fancy}
\fancyhf{}
\renewcommand{\headrulewidth}{0pt}
\lhead{\textcolor{gr}{Physics Letters B 668, 432-437 (2008)}}
\rhead{\textcolor{gr}{ULB-TH/08-20}}
\preprint{\textcolor{gr}{ULB-TH/08-20}}

\title{Instantaneous interaction in massive gravity}


\author{Michael V. Bebronne}
\affiliation{Service de Physique Th\'eorique, Universit\'e Libre de Bruxelles (U.L.B.), CP225, boulevard du Triomphe, B-1050 Bruxelles, Belgium.}
%\thanks{}
%\affiliation{}


\date{\today}

\begin{abstract}
In general relativity, the instantaneous contributions to the gravitational potentials cancel each other in observable, leaving the theory free of physical instantaneous interactions. In models where these subtle cancellations are spoiled by the presence of fields that break Lorentz invariance, physical instantaneous interactions are possible. Such interactions are studied for a model of Lorentz-violating massive gravity.
\end{abstract}

%\keywords{}
%\preprint{ULB-TH/...}
\maketitle

\thispagestyle{fancyplain}

\section{Introduction}

Gravity is the most familiar but also the most mysterious of the four fundamental interactions. Described by Einstein's theory of general relativity (GR), gravity has only been tested in the weak field limit \cite{Will:2005va}. With the development of gravitational wave experiments, it is believed that our understanding of gravity physics will significantly increase in the next decades \cite{Thorne:1995xs}. Although indirect detection of gravitational waves seems to be in agreement with Einstein's theory \cite{Hulse:1994zz,Taylor:1994zz,Kramer:2006nb}, models of gravity with different predictions than GR will be needed in order to study the results of these experiments. Indeed, it is important to understand which deviation from Einstein's theory are possible in order to put constraint on them.

There were many attempts in constructing alternative theories of gravity in the past few years \cite{Milgrom:1983pn,Gregory:2000jc,Dvali:2000hr,Kogan:2000vb,Damour:2002ws,Carroll:2003wy,Arkani-Hamed:2003uy,Bekenstein:2004ne}, motivated by cosmological observations. Although the Standard Model of cosmology describes the whole bulk of cosmological observations with an ever growing precision \cite{Dunkley:2008ie}, the agreement between observations and theory is only possible with the account of otherwise undetected dark components : dark matter ($\Omega_{c} \sim 0.22$) and dark energy ($\Omega_\Lambda \sim 0.74$). Alternative theories of gravity could then help understanding whether this agreement should be consider as a confirmation of GR.

Black holes are perhaps the most interesting objects to constrain alternative models of gravity, since by reconstructing the metric around an astrophysical black hole we should be able to test whether it has the Schwarzschild or Kerr form, and therefore probe GR in the strong field limit. A crucial feature of black holes is the absence of ``hair'' \cite{PhysRevD.5.2941}, which simply means that a black hole is entirely characterized by its mass, charge and angular momentum. This ``no-hair'' theorem is a consequence of the existence of an horizon which divides the space-time in two region: the outside and the inside of the black hole. The causal structure of the metric implies then that no signal with velocity smaller or equal to the velocity of light can escape from the inside of the black hole. In some alternative models of gravity, there are other ``gravitational'' fields beside the metric. It may happen that one of them is responsible for an instantaneous interaction; in this case black holes should in principle have ``hair'' since this instantaneous field could carry information outside of the horizon.

Even in the conventional GR, the gravitational potentials are all instantaneous potentials. But there is no physical instantaneous interaction since the instantaneous contributions cancel in the graviton propagator. An analogous situation exist in classical electrodynamic, where the instantaneous contributions to the potentials $A_0$ and $A_i$ cancel each other in the photon propagator, leaving the theory free of instantaneous interactions \cite{brill:832}. In models where these subtle cancellations are spoiled by the presence of fields that break Lorentz invariance, physical instantaneous interactions are possible. An example of such a model is given by the Lorentz-violating electrodynamics \cite{Dvali:2005nt,Gabadadze:2004iv}. The goal of this Letter is to demonstrate by explicit calculation that a gravitational instantaneous interaction exists in Lorentz-breaking models of massive gravity as well. This interaction is assumed to be responsible for unusual properties of black holes found in these models \cite{Dubovsky:2007zi}, i.e., the presence of ``hair''.

For this purpose, we consider the theory of massive gravity \cite{Dubovsky:2005dw} described by the following action
\begin{eqnarray} \label{eq:MGaction}
\mathcal{S} = \int \dif^4 x \sqrt{- g} \left[ - M_{pl}^2 \mathcal{R} + \Lambda^4 \mathcal{F} \left( Z^{ij} \right) + \mathcal{L}_{\textrm{matter}} \right] .
\end{eqnarray}
The first term in this action is the standard Einstein-Hilbert action and $\mathcal{L}_{\textrm{matter}}$ stands for the minimally coupled ordinary matter. $\Lambda$ is a UV cutoff and $\mathcal{F} \left( Z^{ij} \right)$ is a function of the derivatives of four scalar Goldstone fields $\phi^\mu$ which depends on a single argument $Z^{ij} = X^\gamma W^{ij}$, where $\gamma$ is a free parameter and
\begin{eqnarray*} \label{XVW}
X = \Lambda^{-4} g^{\mu\nu} \partial_\mu \phi^0 \partial_\nu \phi^0 , && V^i = \Lambda^{-4} g^{\mu\nu} \partial_\mu \phi^0 \partial_\nu \phi^i , \nonumber \\
Y^{ij} = \Lambda^{-4} g^{\mu\nu} \partial_\mu \phi^i \partial_\nu \phi^j , && W^{ij} = Y^{ij} - \dfrac{V^i V^j}{X} \, .
\end{eqnarray*}
The ground state is translationally invariant, due to the derivative nature of the Goldstone coupling, and spontaneously breaks the Lorentz symmetry (the explicit form of the ground state is given by expression (\ref{eq:vacuum}) in Sect~\ref{sec:MassGravi}). This allows for the absence of \emph{van Dam-Veltman-Zakharov} discontinuity \cite{vanDam:1970vg,Dubovsky:2004sg} and strong coupling problems \cite{Rubakov:2004eb}.

The particular dependence of the function $\mathcal{F}$ on a single argument $Z^{ij}$ ensures that the model is invariant under the following symmetry
\begin{eqnarray*}
\phi^i \rightarrow \phi^i + \chi^i \left( \phi^0 \right) ,
\end{eqnarray*}
where $\chi^i$ are arbitrary functions of $\phi^0$. This symmetry implies the non-pathological behavior of the perturbations about the vacuum solution \cite{Dubovsky:2004sg}, namely, the absence of ghosts and rapid instabilities. This model is a minimal infrared modification of GR in the sense that it does not contain new light propagating degrees of freedom as compared to Einstein gravity. Indeed, the low-energy spectrum consists only of two massive propagating tensor modes, the graviton acquiring a mass due to the space-time dependent condensates of the four Goldstone fields $\phi^\mu$.

This model posses several interesting and peculiar features (see \cite{Rubakov:2008nh} for a review on massive gravity). First of all, the gravitational interaction between time-dependent sources is described by the standard GR's scalar potentials despite the non-zero mass of the graviton. Due to this feature the model passes the terrestrial and solar system tests even for graviton masses as large as $(10^{15}{\rm cm})^{-1}$. Moreover, for $- 1 < \gamma < 0$ and for $\gamma = 1$, the cosmological perturbations in this model behave identically to the perturbations in GR \cite{Bebronne:2007qh}, so that the model seems to be compatible with cosmological observations.

At last, black holes in this model do have ''hair`` \cite{Dubovsky:2007zi}. The violation of the ''no-hair`` theorem is assumed to be the consequence of a physical instantaneous interaction. The existence of such an interaction is related to the presence in the model (\ref{eq:MGaction}) of a mode with the dispersion relation $k^2 = 0$. This mode could be interpreted as a mode with the dispersion relation $\omega^2 = \nu^2 k^2$ and an infinite velocity $\nu$. It is worth noting that the causality paradoxes associated with superluminal propagation are avoided in this model since the spontaneous breaking of the Lorentz symmetry implies a preferred frame. The aim of this Letter is to demonstrate the existence of this physical instantaneous interaction.

The Letter is organized as follows. In Sect.~\ref{sec:Electro} we review the Lorentz-violating electrodynamic model, which share some features with the model (\ref{eq:MGaction}). The emergence of the instantaneous interaction in Lorentz-breaking electrodynamics is analogous to the emergence of the instantaneous interaction in our massive gravity model. In Sect~\ref{sec:MassGravi} we discuss the solutions to the linearized gravitational field equations over Minkowski background. In Sect~\ref{sec:Doppler} we demonstrate the presence of instantaneous interaction by studying the frequency shift of an electromagnetic radiation. Sect~\ref{sec:Discussion} contains the summary of the results and their discussion.

\section{Lorentz-violating massive electrodynamic} \label{sec:Electro}

As a example of physical instantaneous interaction, consider the Lorentz-violating electrodynamic model \cite{Dvali:2005nt,Gabadadze:2004iv}, described by the following Lagrangian
\begin{eqnarray*}
\mathcal{L} = - \dfrac{1}{4} F_{\mu\nu} F^{\mu\nu} + \dfrac{1}{2} m_0^2 A_0^2 - \dfrac{1}{2} m_1^2 A_i^2 + A_{\mu} J^{\mu} .
\end{eqnarray*}
The Minkowski metric is chosen as $(1,-1,-1,-1)$, and $J^\mu$ is a conserved current $\partial_\mu J^\mu = 0$. The Lorentz-invariant Proca theory of massive electrodynamic is recovered by imposing $m_0 = m_1$. Since this Lagrangian is still invariant under spatial rotations, the following parametrization of the spatial vectors $A_i$ and $J_i$ will simplify the equations of motion
\begin{eqnarray*}
A_i = A_i^T + \partial_i A^L , && J_i = J_i^T + \partial_i J^L ,
\end{eqnarray*}
where $A_i^T$ and $J_i^T$ are transverse. It is important to notice that the transverse and longitudinal currents $J_i^T$ and $\partial_i J^L$ extend over all space, even if $J_i$ is localized \cite{1975clel.book.....J}. Indeed, for all localized currents $J_i$, the transverse current $J_i^T$ possesses an unlocalized contribution which is the opposite of the unlocalized contribution of $\partial_i J^L$ \footnote{The transverse and longitudinal currents $J_i^T$ and $\partial_i J^L$ associated to the localized current $J_i$ can be determined using the transverse and longitudinal projection operator introduced in \cite{brill:832}. By taking an explicit example, it is then possible to show that these currents extend over all space.}.

With these notations, the equations of motion are
\begin{eqnarray*}
\partial_i^2 \left( A_0 - \dot{A}^L \right) - m_0^2 A_0 - J_0 &=& 0 , \\
\partial_i \left[ \partial_0 \left( A_0 - \dot{A}^L \right) - m_1^2 A^L - J^L \right] &=& 0 , \\
\left( \Box + m_1^2 \right) A_i^T + J_i^T &=& 0 ,
\end{eqnarray*}
where over-dot denotes the time derivative, and $\Box = \partial_0^2 - \partial_i^2$. The potential $A_0 - \dot{A}^L$ is invariant under the gauge transformations, and for $m_0 = m_1 = 0$, these equations are precisely those of classical electrodynamics.

The first two equations imply the generalized Proca constraint $m_0^2 \dot{A}_0 = m_1^2 \partial_i^2 A^L$. Let us concentrate on the model defined by $m_1 > 0$ and $m_1 \neq m_0$. Then, $m_0$ has to be zero to guarantee the absence of ghosts and rapid instabilities. Since we are interested in solutions which do not grow at infinity, the Proca constraint implies that $A^L = 0$ and the field equations are equivalent to
\begin{eqnarray*}
A_0 = \dfrac{J_0}{\partial_i^2} , && \left( \Box + m_1^2 \right) A_i^T = - J_i^T .
\end{eqnarray*}
If $J_0 \neq 0$, it is obvious from these equations that $A_0$ is an instantaneous potential, since $J_0$ and $A_0$ have the same time dependence. In fact, $A_i^T$ is also an instantaneous potential. Indeed, the transverse current $J_i^T$ extends over all space and implies a non-vanishing $A_i^T$ in all space. As a consequence, for all localized currents $J_i$, $A_i^T$ is an instantaneous potential \footnote{There is only one possibility for $A_i^T$ and $A_0$ not to be instantaneous potentials : $A_i^T$ will not be an instantaneous potential if the localized source is transverse $J_i = J_i^T$ while $A_0$ will not be an instantaneous potential if the localized source is longitudinal $J_i = \partial_i J^L$. }.

Using the current conservation, the equation for $A_0$ can be written $\Box A_0 = \dot{J}^L - J_0$. Hence, the solutions to the field equations are
\begin{eqnarray*}
A_0 &=& \int \dif^4 x^\prime G^{+} \left( t, x , t^\prime, x^\prime \right) \left[ \dot{J}^L \left( t^\prime, x^\prime \right) - J_0 \left( t^\prime, x^\prime \right) \right] , \\
A_i^T &=& - \int \dif^4 x^\prime G_m^{+} \left( t, x , t^\prime, x^\prime \right) J_i^T \left( t^\prime, x^\prime \right) ,
\end{eqnarray*}
where $G^{+} \left( t, x , t^\prime, x^\prime \right)$ and $G_m^{+} \left( t, x , t^\prime, x^\prime \right)$ are the retarded Green functions of the d'Alembert and Klein-Gordon equations
\begin{eqnarray*}
\Box G^{+} \left( t, x , t^\prime, x^\prime \right) &=& \delta^4 \left( x - x^\prime \right) , \\
\left( \Box + m^2 \right) G_m^{+} \left( t, x , t^\prime, x^\prime \right) &=& \delta^4 \left( x - x^\prime \right) ,
\end{eqnarray*}
respectively.

In classical electrodynamics, the instantaneous contributions to the potentials $A_0$ and $A_i^T$ cancel each other in observables such as the electric and magnetic fields \cite{brill:832}. There is then no instantaneous interaction. This is no more true in the Lorentz-violating model, because the mass $m_1$ modifies the dispersion relation of the transverse modes $A^T_i$ without affecting the dispersion relation of $A_0$. Therefore, the instantaneous contributions of the potentials do not cancel each other anymore, giving rise to a physical instantaneous interaction. This can be illustrated by looking at the electric field $E_i \equiv F_{0i}$, which reads
\begin{eqnarray} \label{eq:ElectricField}
E_i &=& \int \dif^4 x^\prime G^{+} \left( t, x , t^\prime, x^\prime \right) \left[ \dfrac{\partial J_0 \left( t^\prime, x^\prime \right)}{\partial x^{i \prime}} - \dfrac{\partial J_i \left( t^\prime, x^\prime \right)}{\partial t^\prime} \right] \nonumber \\
%%%
& & - \int \dif^4 x^\prime \Delta G^{+} \left( t, x , t^\prime, x^\prime \right) \dfrac{\partial J_i^T \left( t^\prime, x^\prime \right)}{\partial t^\prime}
\end{eqnarray}
where $\Delta G^{+} \left( t, x , t^\prime, x^\prime \right) \equiv G_m^{+} \left( t, x , t^\prime, x^\prime \right) - G^{+} \left( t, x , t^\prime, x^\prime \right)$ is proportional to $m_1$. The first integral in (\ref{eq:ElectricField}) is the retarded electric field of classical electrodynamics. This term vanishes outside of the light-cone of the source, since it involves only localized current $J_\mu$, whose effects are retarded because of the retarded Green function $G^{+}$. The second integral is a contribution specific to the massive case, which does not vanish outside of the light-cone of the source since it involves the unlocalized transverse current $J_i^T$. Therefore, this second integral extends over all space and represents an instantaneous contribution to the electric field.

It is worth noting that the presence of this physical instantaneous interaction is related to the breaking of the Lorentz invariance. Without this symmetry breaking, the modifications to the dispersion relation of $A_0$ and $A^T_i$ would be proportional and the instantaneous contributions of these potentials would cancel each other in the expression (\ref{eq:ElectricField}). The discussion of physical consequences of instantaneous interactions in massive electrodynamics can be found in \cite{Dvali:2005nt}.

\section{Lorentz-violating massive gravity} \label{sec:MassGravi}

The emergence of a physical instantaneous interaction in the massive gravity model described by (\ref{eq:MGaction}) is also a consequence of the non-cancellation of two unlocalized contributions, whereas those contributions cancel in GR. Therefore, the same approach as the one used in the previous section will be useful to understand how a physical instantaneous interaction appears in this model.

Gravity is described by a full non-linear theory. Therefore, small perturbations about the flat vacuum solution will be consider. For an empty space, the energy-momentum of the usual matter is zero, $\mathcal{T}_{\mu\nu} = 0$, and the following vacuum configuration is solution to the Einstein equations of motion
\begin{eqnarray} \label{eq:vacuum}
\begin{array}{ccccc}
g_{\mu\nu} = \eta_{\mu\nu} &,& \phi^0 = a \Lambda^2 t &,& \phi^i = b \Lambda^2 x^i \, .
\end{array}
\end{eqnarray}
Here $a$ and $b$ are two constant whose value is set by the requirement that the energy-momentum tensor of the four Goldstone fields vanishes in Minkowski space-time. As mentioned in the introduction, this Goldstone vacuum solution is not Lorentz-invariant. Therefore, there is a spontaneous breaking of the Lorentz symmetry in this massive gravity model.

A small perturbation $\delta \mathcal{T}_{\mu\nu}$ of the matter energy-momentum tensor will produce metric $\delta g_{\mu\nu}$ and Goldstone $\delta \phi^\mu$ perturbations about the vacuum solution (\ref{eq:vacuum}), just as a non-vanishing current $J_\mu$ is responsible for the electromagnetic potentials $A_\mu$. It is convenient to parameterize the energy-momentum tensor of the source in the following way,
\begin{eqnarray*}
\delta \mathcal{T}_{\mu\nu} = \left( \delta \rho + \delta p \right) v_{\mu} v_{\nu} - \eta_{\mu\nu} \delta p + \left( v_{\mu} \delta q_{\nu} + v_{\nu} \delta q_{\mu} \right) + \delta \pi_{\mu\nu} , \label{energy-tensor-pert}
\end{eqnarray*}
where $\delta \rho$ and $\delta p$ are the matter density and pressure measured by a comoving observer of velocity $v_\mu$, $\delta q_\mu$ is the energy flux perpendicular to $v_\mu$ ($v^\mu \delta q_\mu = 0$) and $\delta \pi_{\mu\nu}$ is the anisotropic pressure tensor ($v^\mu \delta \pi_{\mu\nu} = \delta \pi_{\mu}^\mu = 0$). The velocity $v_\mu$ of the observer obeys the geodesic equation $v^\nu \partial_\nu v^\mu = 0$, with $v^\mu v_\mu = 1$. This implies that the affine parameter of the observer can be chosen such that $v_\mu = \left( 1, 0, 0, 0 \right)$. As a consequence, $\delta q_0 = \delta \pi_{0\nu} = 0$ and
\begin{eqnarray*}
\delta \mathcal{T}_{00} = \delta \rho , & \delta \mathcal{T}_{0i} = \delta q_i , & \delta \mathcal{T}_{ij} = \delta_{ij} \delta p + \delta \pi_{ij} .
\end{eqnarray*}
The energy flux $\delta q_i$ and the anisotropic pressure $\delta \pi_{ij}$ can be parameterized in the following way,
\begin{eqnarray*}
\delta q_i &=& \zeta_i + \partial_i \zeta , \\
%%%
\delta \pi_{ij} &=& \left( 3 \partial_i \partial_j - \delta_{ij} \partial_k^2 \right) \pi + \partial_i \pi_j + \partial_j \pi_i + \pi_{ij} ,
\end{eqnarray*}
where the vector perturbations $\zeta_i$ and $\pi_i$ are transverse, while the tensor perturbation $\pi_{ij}$ is transverse and traceless. As in electrodynamics, it is important to notice that the transverse and longitudinal parts of $\delta \mathcal{T}_{\mu\nu}$ extend over the whole space, even if $\delta \mathcal{T}_{\mu\nu}$ is localized. In particular, $\pi_{ij}$ has a non-vanishing value everywhere in space for all localized energy-momentum tensors $\delta \mathcal{T}_{\mu\nu}$.

With these notations, the energy-momentum conservation reads
\begin{eqnarray*}
\dot{\delta \rho} = \partial_i^2 \zeta , & \dot{\zeta} = \delta p + 2 \partial_i^2 \pi , & \dot{\zeta_i} = \partial_j^2 \pi_i .
\end{eqnarray*}
Hence, the ten independent components of the energy-momentum tensor are expressed trough four gauge-invariant scalars $\delta\rho$, $\delta p$, $\zeta$ and $\pi$, four vector degrees of freedom in the form of two transverse gauge-invariant vectors $\zeta_i$ and $\pi_i$, and two tensor degrees of freedom trough the transverse and traceless gauge-invariant tensor $\pi_{ij}$.

As mention earlier, the presence of the source $\delta \mathcal{T}_{\mu\nu}$ implies a modification of the solution (\ref{eq:vacuum}) for the metric and Goldstone fields, i.e., the metric and Goldstone perturbations. We parameterize the metric perturbations as following
\begin{eqnarray*}
& \begin{array}{cc}
\delta g_{00} = 2 \varphi , & \delta g_{0i} = S_i + \partial_i B ,
\end{array} & \\
& \delta g_{ij} = 2 \Psi \delta_{ij} - 2 \partial_i \partial_j E - \partial_i F_j - \partial_j F_i + h_{ij} & .
\end{eqnarray*}
A similar parametrization can be used for the perturbations of the Goldstone fields,
\begin{eqnarray*}
\delta\phi^0 = a \Lambda^2 \xi^0 &,& \delta \phi^i = b \Lambda^2 \left( \xi_i + \partial_i \xi \right) .
\end{eqnarray*}
The vector perturbations $S_i$, $F_i$ and $\xi_i$ are transverse, while the tensor perturbation $h_{ij}$ is transverse and traceless. Finally, it will be useful to introduce gauge-invariant fields. One vector and two scalar perturbations are gauge degrees of freedom. As a consequence, there is only two gauge-invariant vector fields
\begin{eqnarray*}
\varpi_i = S_i + \dot{F}_i , & & \sigma_i = \xi_i - F_i
\end{eqnarray*}
and four scalar gauge-invariant fields
\begin{eqnarray*}
\Phi = \varphi - \left( \ddot{E} + \dot{B} \right) , & & \Xi = \xi -  E, \\
\Xi^0 = \xi^0 - \dot{E} - B & &
\end{eqnarray*}
and $\Psi$. The tensor perturbation $h_{ij}$ is also gauge invariant.

With the account of all these notations, the linearized Einstein equations split into scalar, vector and tensor equations. These tree sectors could then be consider separately.

\subsection{The scalar perturbations}

The behavior of the four scalar gauge-invariant perturbations is governed by four Einstein equations (see Ref.\cite{Bebronne:2007qh} for details). The solutions of these equations, which does not grow at spatial infinity, reads
\begin{eqnarray*}
\Psi = \Psi_E + \Psi_0 \left( x^i \right) , & & \Phi = \Phi_E + \left( 1 - \dfrac{1}{\gamma} \right) \Psi_0 \left( x^i \right) ,
\end{eqnarray*}
where $\Psi_0 \left( x^i \right)$ is an arbitrary function of the space coordinates, and $\Psi_E$ and $\Phi_E$ are the scalar potentials of GR ($\Phi_E$ is nothing else than the Newtonian potential)
\begin{eqnarray*}
\Psi_{E} = \dfrac{\delta \rho}{2 M_{pl}^{2} \partial_i^2} , & & \Phi_{E} = \dfrac{1}{M_{pl}^{2}} \left[ \dfrac{\delta \rho}{2 \partial_i^2} - 3 \pi \right] .
\end{eqnarray*}
It is obvious from these two equations that $\Psi_E$ and $\Phi_E$ are two instantaneous potentials, such as the Coulomb potential $A_0 - \dot{A}^L$ of electrodynamics. In the absence of anisotropic pressure, $\pi = 0$, GR predicts $\Phi = \Psi$. This relation is not satisfied in this massive gravity model because of the appearance of the function $\Psi_0$, which value is fixed by initial conditions \cite{Dubovsky:2005dw}. Hence, if $\Psi_0 \neq 0$ the gravitational potential $\Psi$ created by a massive source is not the potential $\Phi$ responsible for the geodesic motion of an observer around this source. Therefore, if we neglect this arbitrary function there is no difference in the scalar sector between GR and the model given by (\ref{eq:MGaction}).

With the account of the energy-momentum conservation, these two instantaneous potentials can be expressed trough
\begin{eqnarray*}
\Box \Psi_E &=& \dfrac{1}{2 M_{pl}^2} \left( \dot{\zeta} - \delta \rho \right) , \\
\Box \Phi_E &=& \dfrac{1}{M_{pl}^2} \left( 2 \dot{\zeta} - 3 \ddot{\pi} - \dfrac{\delta \rho + 3 \delta p}{2} \right) .
\end{eqnarray*}
Then, using the Green function of the d'Alembert equation,
\begin{eqnarray} \label{eq:Scalars}
\Psi_E &=& \int \dfrac{\dif^4 x^\prime}{2 M_{pl}^2} G^{+} \left( t, x, t^\prime, x^\prime \right) \left[ \dot{\zeta} \left( t^\prime, x^\prime \right) - \delta \rho \left( t^\prime, x^\prime \right) \right] , \nonumber \\
\Phi_E &=& \int \dfrac{\dif^4 x^\prime}{M_{pl}^2} G^{+} \left( t, x, t^\prime, x^\prime \right) \left[ 2 \dot{\zeta} \left( t^\prime, x^\prime \right) \right. \nonumber \\
& & \left. - 3 \ddot{\pi} \left( t^\prime, x^\prime \right) - \dfrac{\delta \rho \left( t^\prime, x^\prime \right) + 3 \delta p \left( t^\prime, x^\prime \right)}{2} \right] .
\end{eqnarray}

\subsection{The vector perturbations}

It follows from the vector equations that $\varpi_{i}$ is an instantaneous potential, equal to its value in GR, $\varpi_{i} = \varpi^E_{i}$ with
\begin{eqnarray*}
\varpi^{E}_{i} = \dfrac{2}{M_{pl}^{2} \partial_j^2} \zeta_{i} .
\end{eqnarray*}
Therefore, there is no difference in the vector sector between GR and the model of massive gravity given by (\ref{eq:MGaction}). Using the energy-momentum conservation, the equation for $\varpi_{i}$ can be written as $\Box \varpi_{i} = 2 M_{pl}^{-2} \left( \dot{\pi}_{i} - \zeta_{i} \right)$, which gives the following solution
\begin{eqnarray} \label{eq:Vector}
\varpi_{i} = 2 \int \dfrac{\textrm{d}^4 x^\prime}{M_{pl}^2} G^{\pm} \left( t , x , t^\prime , x^\prime \right) \left[ \dot{\pi}_{i} \left( t^\prime , x^\prime \right) - \zeta_{i} \left( t^\prime , x^\prime \right) \right] .
\end{eqnarray}

\subsection{The tensor perturbations}

The equation for the massive gravitational waves is
\begin{eqnarray} \label{eq:Graviton}
0 &=& \left( \Box + m_2^2 \right) h_{ij} + \dfrac{2 \pi_{ij}}{M_{pl}^{2}},
\end{eqnarray}
where $m_2 \propto \Lambda^2 / M_{pl}$ is the mass of the graviton (see Ref.\cite{Bebronne:2007qh} for a precise expression of $m_2$ in terms of $\mathcal{F} \left( Z^{ij} \right)$ and its derivatives). The tensor potential $h_{ij}$ is instantaneous for the same reason as the potential $A^T_i$ of electrodynamics. Indeed, for all physical situations $\delta \mathcal{T}_{ij}$ is localized in space. But its transverse and traceless part $\pi_{ij}$ extend over the whole space, implying a non-vanishing tensor potential $h_{ij}$, everywhere in space.

Using the Green function of the Klein-Gordon equation, the tensor potential is
\begin{eqnarray} \label{eq:Tensor}
h_{ij} = - 2 \int \dfrac{\dif^4 x^\prime}{M_{pl}^2} G_m^{+} \left( t, x, t^\prime, x^\prime \right) \pi_{ij} \left( t^\prime, x^\prime \right) .
\end{eqnarray}
In this equation, we neglect the tensor modes which are solutions to the homogeneous equation $\left( \Box + m_2^2 \right) h_{ij} = 0$, since these modes are present in the vacuum and do not play any role in the interaction between gravitational sources.

At this point, let us stress that the only difference in the linearized theory between GR and the model (\ref{eq:MGaction}) lays in the tensor sector. The modification of the dispersion relation of $h_{ij}$ induced by $m_2$ is identical to the modifications of the dispersion relation of the vector field $A_i^{T}$ of electrodynamics induced by $m_1$. Since there is no modifications of dispersion relations in the other sectors of the theory, due to the breaking of the Lorentz invariance, the instantaneous contributions of the potentials will not cancel themselves in the graviton propagator. Therefore, it is necessary to search for observable involving $h_{ij}$ in order to see a physical instantaneous interaction.

\section{Frequency shift} \label{sec:Doppler}

To illustrate the physical instantaneous interaction present in our model of massive gravity, consider the measured energy of a light beam. Let $v^\mu$ be the four-velocity of an observer, with $v^\mu v_\mu = 1$, and $u^\mu$ the vector tangent to the light's geodesic, with $u^\mu u_\mu = 0$. The light's momentum is defined by $p^\mu = \omega_0 u^\mu$, where $\omega_0$ is a constant, and the frequency measured by the observer is given by
\begin{eqnarray} \label{eq:energy}
\omega = v_\mu p^\mu .
\end{eqnarray}

In the vacuum $\mathcal{T}_{\mu\nu} = 0$ and the solution (\ref{eq:vacuum}) holds. Then, the affine parameter $\tau_{ob}$ of the observer can be chosen such that $v^\mu = \left( 1, 0, 0, 0 \right)$, while the affine parameters $\tau_{ph}$ of the light's geodesic can be chosen such that $u^\mu = \left( 1, n^i \right)$ with $n_i^2 = 1$. Therefore, the frequency measured by the observer is $\omega = \omega_0$.

If there is a small extra source for gravity, $\mathcal{T}_{\mu\nu} = \delta \mathcal{T}_{\mu\nu}$, the solution (\ref{eq:vacuum}) is slightly modified. Because of the geodesic equations, the perturbations of the metric imply perturbations $\delta v^\mu$ of the four-velocity of the observer and perturbations $\delta u^\mu$ of the vector tangent to the light's geodesic. Therefore, $\delta v^\mu$ and $\delta u^\mu$ are given in terms of the metric perturbations and there is a shift of the frequency measured $\delta \omega = \omega \left( \delta v_\mu u^\mu + v_\mu \delta u^\mu \right)$. At the linearized level, equation (\ref{eq:energy}) gives
\begin{eqnarray} \label{eq:shift}
\dfrac{\delta \omega}{\omega} &=& \int \dif \tau_{ph} \left[ \dot{\Psi} - n^i \partial_i \Phi + n^i n^j \left( \dfrac{1}{2} \dot{h}_{ij} - \partial_{i} \varpi_j \right) \right] \nonumber \\
& & + \int \dif \tau_{ob} \, n^i \partial_i \Phi .
\end{eqnarray}
In the Friedmann-Robertson-Walker background, this relation is known as the \emph{Sachs-Wolfe} effect. From (\ref{eq:shift}) it is obvious that any difference in the tensor sector of the theory compared to GR will lead to a modification of the spectral shift measured.

\subsection{Frequency shift in GR}

In GR, the scalar and vector potentials are given by relations (\ref{eq:Scalars}) and (\ref{eq:Vector}). The tensor potential $h^{E}_{ij}$ is a solution to equation (\ref{eq:Graviton}), with $m_2^2 = 0$. Therefore, this potential is given by (\ref{eq:Tensor}) where the Green function $G^{+}_m$ is replaced by the Green function $G^{+}$, since $G^{+}_m \rightarrow G^{+}$ when $m_2 \rightarrow 0$. With the account of these tree relations, equation (\ref{eq:shift}) gives the frequency shift measured by the observer in GR
\begin{widetext}
\begin{eqnarray*} \label{eq:ShiftGR}
\dfrac{\delta \omega^{E} \left( t, x \right)}{\omega} &=& \int \dfrac{\textrm{d}^4 x^\prime}{M_{pl}^2} G^{+} \left( t, x, t^\prime, x^\prime \right) \left[ \dfrac{\delta \mathcal{T}_{kk} \left( t^\prime , x^\prime \right) - \delta \mathcal{T}_{00} \left( t^\prime , x^\prime \right)}{2} - n^i n^j \delta \mathcal{T}_{ij} \left( t^\prime , x^\prime \right) \right] \nonumber \\
%%%
& & - \int \dfrac{\dif \tau_{ob}}{2 M_{pl}^2} \int \textrm{d}^4 x^\prime G^{+} \left( t, x, t^\prime, x^\prime \right) n^i \dfrac{\partial}{\partial x^{\prime i}} \left[ \delta \mathcal{T}_{00} \left( t^\prime , x^\prime \right) + \delta \mathcal{T}_{kk} \left( t^\prime , x^\prime \right) \right] \nonumber \\
%%%
& & + n^i n^j \int \dfrac{\dif \tau_{ph}}{M_{pl}^2} \int \textrm{d}^4 x^\prime G^{+} \left( t, x, t^\prime, x^\prime \right) \dfrac{\partial}{\partial x^{\prime i}} \left[ n^k \left( \delta \mathcal{T}_{jk} \left( t^\prime , x^\prime \right) + \delta_{jk} \delta \mathcal{T}_{00} \left( t^\prime , x^\prime \right) \right) + 2 \delta \mathcal{T}_{0j} \left( t^\prime , x^\prime \right) \right] .
\end{eqnarray*}
\end{widetext}
The instantaneous contributions to the potentials cancel each others in this relation. Indeed, this expression involves only localized sources, i.e., the components of the localized energy-momentum tensor $\delta \mathcal{T}_{\mu\nu}$. The effects of these sources are retarded because of the retarded Green function $G^{+}$. As a consequence, in GR the frequency shift cancels outside of the light cone of the localized source, and there are no instantaneous interactions.

\subsection{Frequency shift in massive gravity}

The situations is different in the massive gravity model described by the action (\ref{eq:MGaction}), where $m_2^2 \neq 0$. Indeed, the scalar and vector potentials are still given by relations (\ref{eq:Scalars}) and (\ref{eq:Vector}). But the tensor potential is given by (\ref{eq:Tensor}), with $m_2^2 \neq 0$. Therefore, the spectral shift measured by an observer of velocity $v^\mu$ is expressed trough
\begin{eqnarray} \label{eq:ShiftMG}
\dfrac{\delta \omega}{\omega} = \dfrac{\delta \omega^E}{\omega} + \int \dif \tau_{ph} \frac{n^i n^j}{2} \dot{h}^\Delta_{ij} ,
\end{eqnarray}
where $h^\Delta_{ij} = h_{ij} - h^{E}_{ij}$. $\Psi_0$ has been neglected in this expression since this unknown function does not play any role in the interaction between time-dependent sources. Hence, there is only one extra term in relation (\ref{eq:ShiftMG}) as compared to GR. This term extends over the whole space and is responsible for an instantaneous shift of frequency. Therefore, the shift is observable outside of the light cone of the source. At this point, let us stress that this extra term has the same shape as the extra electric field (\ref{eq:ElectricField}) obtained in the Lorentz-violating electrodynamic model discussed before, since these two extra terms involve the same function $\Delta G^{+}$ with sources which extend over the whole space
\begin{eqnarray} \label{eq:DeltaTensor}
h^\Delta_{ij} \left( t, x \right) = - 2 \int \dfrac{\textrm{d}^4 x^\prime}{M_{pl}^2} \Delta G^{+} \left( t, x, t^\prime, x^\prime \right) \pi_{ij} \left( t^\prime , x^\prime \right) .
\end{eqnarray}

To demonstrate that this term is responsible for an instantaneous shift of frequency, consider the derivative of the spectral shift with respect to the affine parameters $\tau_{ph}$ of the light geodesic
\begin{eqnarray} \label{eq:derivatives_shift}
\textbf{a}_\omega \equiv \dfrac{\dif}{\dif \tau_{ph}} \left( \dfrac{\delta \omega}{\omega} - \dfrac{\delta \omega^{E}}{\omega} \right) .
\end{eqnarray}
This quantity will now be determined for a concrete example of gravitational source.

\subsection{Instantaneous interaction : an example}

As an explicit example of physical instantaneous interaction, suppose we have a source which appears at $t = 0$, described by the following energy-momentum tensor
\begin{eqnarray}
\delta \mathcal{T}_{00} &=& 2 \mu_{ij} t^2 \Theta \left( t \right) \partial_i \partial_j \delta^3 \left( x - x_s \right) , \nonumber \\
\delta \mathcal{T}_{0i} &=& 4 \mu_{ij} t \Theta \left( t \right) \partial_j \delta^3 \left( x - x_s \right) , \nonumber \\
\delta \mathcal{T}_{ij} &=& 4 \mu_{ij} \Theta \left( t \right) \delta^3 \left( x - x_s \right) , \label{eq:Sources}
\end{eqnarray}
where $x_s = \left( 0, 0, d \right)$. The presence of the Heaviside function in these expressions guarantees that $\delta \mathcal{T}_{\mu\nu} = 0$ before $t = 0$. Let the observer be located at $x^i = 0$ and the light's geodesic be along the $x$ direction, with $n^i = - \delta^i_1$. Since the distance between the observer and the source is $d$, and since the speed of light is one in our units $c = 1$, in GR the observer will not measure any spectral shift before $t = d$.

In the model (\ref{eq:MGaction}), there is an instantaneous interaction and (\ref{eq:derivatives_shift}) is not zero for $t < d$. To determine the frequency shift, the integral (\ref{eq:DeltaTensor}) has to be evaluated for the source (\ref{eq:Sources}). If the source is chosen such that $\mu_{ij} = \delta_{2i} \delta_{2j}$, the only relevant component of the transverse and traceless tensor $\pi_{ij}$ is given by
\begin{eqnarray*}
\pi_{11} = 2 \Theta \left( t \right) \left( \dfrac{\partial_x^2 \partial_y^2}{\partial_i^2 \partial_j^2} - \dfrac{\partial_z^2}{\partial_i^2} \right) \delta^3 \left( x - x_s \right) .
\end{eqnarray*}
Hence, using relations (\ref{eq:ShiftMG}) and (\ref{eq:DeltaTensor}), the derivative of the spectral shift with respect to the affine parameters of the light geodesic is expressed trough
\begin{eqnarray*}
\textbf{a}_\omega \left( t, x \right) &=& - \dfrac{2}{M_{pl}^2} \left( \dfrac{\partial_x^{2} \partial_y^{2}}{\partial_i^{2} \partial_j^{2}} - \dfrac{\partial_z^{2}}{\partial_i^{2}} \right) \Delta G^{+} \left( t, x, 0, x_s \right) .
\end{eqnarray*}
Since $\Delta G^{+}$ involves the Bessel function $\mathcal{J}_{1} \left( m_2 x \right)$ divided by its argument and since $m_2 x \ll 1$, the following Taylor expansion can be used to find an approximation of this last expression
\begin{eqnarray*}
\dfrac{\mathcal{J}_{1} \left( m_2 x \right)}{m_2 x} = \dfrac{1}{2} - \dfrac{m_2^2 x^2}{16} + \mathcal{O} \left( m_2^3 x^3 \right) .
\end{eqnarray*}
With the account of all these relations, the measured variation of the frequency, caused by the source located at distance $d$ of the observer, is simply given for $t < d$ (outside of the light cone of the source) by
\begin{eqnarray*}
\textbf{a}_\omega \left( t , 0 \right) = - \dfrac{m_2^4 \, d}{64 \pi M_{pl}^2} t + \mathcal{O} \left( m_2^5 \right) , & 0 \le t < d .
\end{eqnarray*}
This last expression clearly shows a measurable instantaneous interaction, proportional to the mass of the graviton to the fourth power.

\section{Discussion} \label{sec:Discussion}

Physical instantaneous interactions are present in the massive gravity model (\ref{eq:MGaction}). The existence of such interactions is related to the spontaneously breaking of the Lorentz-symmetry induced by space-time dependent vacuum expectation values of the four Goldstone fields $\phi^\mu$. This symmetry breaking allows for modifications of the dispersion relation of the tensor modes as compared to GR without affecting the dispersion relations of other fields. As a consequence, the instantaneous contributions to the potentials do not cancel in the graviton propagator, unlike in GR. This is analogous to the situation in the Lorentz-violating electrodynamics, where the instantaneous contributions to the potentials do not cancel in the photon propagator.

It has been demonstrated here that a gravitational source localized in space is responsible for an instantaneous frequency shift of light beams seen by an observer. It is worth noting that, at the linearized level, the amplitude of this instantaneous spectral shift is proportional to the graviton mass to the fourth power, and therefore is the instantaneous interaction very small as compared to the usual retarded interaction between gravitational sources. Regardless of the strength, there is no causal paradox associated with superluminal propagation in this model since the breaking of the Lorentz symmetry implies a preferred frame.

The action (\ref{eq:MGaction}) is a low-energy effective action. One should then expect corrections containing higher-derivative terms to be present in the massive gravity model. These corrections could not provide modification of the dispersion relations of the scalar and vector fields proportional to $m_2^2$. Therefore, the instantaneous contributions coming from the mass of the graviton cannot be cancel by higher-derivative terms, and the inclusion of such terms will not affect the conclusion of this work.

Since the appearance of the physical instantaneous interaction is related to the breaking of the Lorentz-symmetry, it will not be surprising if the presence of this interaction is a generic feature of model of gravity where Lorentz-invariance is broken (like in other massive gravity \cite{Rubakov:2004eb} or bigravity \cite{Berezhiani:2007zf,Blas:2007zz} theories).

Finally, the presence of this physical instantaneous interaction is supposed to be responsible for the violation of the black hole ''no-hair`` theorem present in this model. Indeed, it seems reasonable to assume that a instantaneous interaction could carry informations through the black hole horizon, while interactions which propagate at finite velocities $v \le c$ cannot. It is worth noting that this instantaneous interaction should also allow to look behind the cosmological horizon, since the cosmological horizon is somehow similar to the black hole horizon. It would be interesting to study this last issue in more detail.

\begin{acknowledgments}
I am grateful to Peter Tinyakov for stimulating discussions and for his useful contributions to this manuscript. This work is supported by the Belgian \emph{Fond pour la Formation \`a la Recherche dans l'Industrie
et dans l'Agronomie (FRIA)}.
\end{acknowledgments}

\end{document}